\documentstyle[11pt,aaspp4]{article}

\slugcomment{submitted to Astrophysical Journal Letters}

\lefthead{Newberg and Yanny}
\righthead{Absence of gaps}

\begin{document}

\title{An Absence of Gaps in the Main Sequence Population of Field Stars}

\author{Heidi Jo Newberg and Brian Yanny}
\affil{Fermi National Accelerator Laboratory, Box 500, Batavia, IL 60510}
\centerline{\it heidi@fnal.gov, yanny@fnal.gov}


\begin{abstract}

Using high precision parallaxes from the Hipparcos catalog, we
construct H-R diagrams for two samples of bright stars.  The first is
a magnitude-limited sample that is over $90\%$ complete and uses
uniform photometry from the Catalog of WBVR Magnitudes of Northern Sky
Bright Stars ($\delta > -14^\circ$).  This sample shows a smooth 
distribution of stars along the main sequence, with no detectable gaps.  
The second contains all of the stars closer than 100 parsecs in the 
Hipparcos catalog with $\delta < -12^\circ$.  Uniform spectroscopy 
from the Michigan Spectral Survey shows that some stars which appear 
on the main sequence in the H-R diagram, particularly those in the 
$0.2 < B-V < 0.3$ region that has been labeled the B\"ohm-Vitense gap, 
are classified as giants by the MK system of spectral classification.  
Other gaps that have been identified in the main sequence are also 
affected by such classification criteria.  This analysis casts doubt
on the existence of the B\"ohm-Vitense gap, which is thought 
to result from the sudden onset of convection in stars.  The standard 
identification of main sequence stars with luminosity class V, and giants 
with luminosity class III, must be reconsidered for some spectral types. 
The true nature of the stars that lie on the main sequence in the H-R diagram,
but which do not have luminosity class V designations, remains to be
investigated.

\end{abstract}




%

\section{Introduction}

Classical heat transport theory applied to stars suggests that fully
stably-stratified radiative transport occurs for unevolved stars with
hotter effective temperatures, while cooler stars have increasingly
deep convective envelopes (e.g. \cite{s58}).  Astronomers have
searched extensively for clear observational evidence of a
transition region between those with radiative and those with
convective upper layers.

In particular, \cite{bv70} suggested that stars with convective
atmospheres would be approximately 0.08 magnitudes redder in $B-V$
than stars with radiative upper layers for models with the same
$T_{\rm eff}$, at least in the radiative-convective transition range.
This implies that a population of main sequence stars with a
continuous range of effective temperatures could exhibit a 0.08
magnitude jump in $B-V$ at some critical transition effective
temperature $T_{\rm tr}$.  This theoretical phenomenon has been
referred to as the ``the abrupt onset of convection'' and the apparent
gap is referred to as the B\"ohm-Vitense gap.

Although a gap was apparent in the local population of
main sequence field stars as far back as 1953 (\cite{jm53}),
\cite{bv70} and \cite{bvc74} first associated
the gap with the onset of convection.  Observations of this 
gap in the field population (\cite{mhj53}, B\"ohm-Vitense \& Canterna 1974,
 Jasniewicz 1984)
have generally been obtained by using spectroscopically
determined luminosity classes to select only main sequence (luminosity class V)
objects.  For field stars the gap is typically in the
$0.2 < B-V < 0.3$ range, which is consistent with the 
prediction of \cite{bv70}.

One would imagine that such a gap, as obvious as it is in the field
population, would be easily detected in open clusters.
Observations of clusters, however,
have produced ambiguous results.  For example, \cite{m56} finds no
gaps in the Pleiades, while \cite{j84}, using stricter selection
criteria, does note a gap.  The latter author observes the position
in color of the blue edge
of the B\"ohm-Vitense gap to apparently differ with cluster age.
\cite{cpc79} and \cite{kf91} note that there are many different gaps,
while \cite{hfmr93} find hints of a gap, but conclude that it may not
be significant.  Even \cite{bvc74} do not see a gap in all clusters.
Evidently, the statistical significance, location in color space, and
density depression of the gaps appears to change from cluster to
cluster.

One might account for the differing positions of the gaps in different
clusters if rapid rotation of the stars in some clusters
systematically retarded the onset of convection 
(B\"ohm-Vitense \& Canterna 1974).
\cite{bv82} suggested that differing rotations, and thus different
critical temperatures for the onset of convection, would cause a
spread in a plot of $T_{\rm eff}$ vs. $B-V$.  Based on a sample of
about 50 stars, the paper finds evidence for two branches of late A
and early F stars.  Here, $T_{\rm eff}$ was determined from
ultraviolet observations which probe higher atmospheric layers which
were claimed by \cite{bv82} to be a better predictor of a star's
$T_{\rm eff}$ than the $B-V$ color index.  Recently, however,
\cite{sl97b} obtained more UV observations of A and F stars and find no
evidence for a bifurcation in the ($B-V$, $T_{\rm eff}$) plane, and
suggest that there may be no such phenomenon.

An alternate explanation of the apparent gaps is
suggested by Figure 4 of \cite{ny97}.  Here one can see two depletions
in the distribution of over 3500 luminosity class IV and V field
stars: one at $B-V \approx 0.0$, and one at $B-V \approx 0.3$.  These
depletions are depicted to coincide with a relative excess of stars
identified as luminosity class III.  In this letter, we use the recent
publication of the full Hipparcos data set (\cite{p97}) to
investigate whether
systematics of the classification could be responsible for the gaps
found by other authors.

\section {Flux-limited sample -- WBVR photometry} 

The Catalog of WBVR Magnitudes of Northern Sky Bright Stars
(\cite{kmz96}, \cite{km94}) provides uniform and highly accurate
photometry for $95\%$ of the stars with $V < 7$ and $\delta > -14\deg$.  
Of the 9253 WBVR stars which match these criteria, 8896 stars have Hipparcos
proper motions.  The resulting catalog is a ($90\%$ complete) flux-limited
sample of field stars with $M_V$ derived from parallaxes.  The H-R diagram for 
this set of stars (Fig. 1a) does not
show any gaps or under-densities in the $0.2 < B-V < 0.3$ region of the
main sequence, but rather shows a smooth distribution of stars.

Over half of the stars (5717) have luminosity classes assigned to
them.  The H-R diagrams for luminosity class III, IV, and V stars for
which luminosity classes exist are plotted in Fig 1 (b, c, d).  Note
that in Figure 1d there are depletions in the stellar density at
$B-V \approx 0.3$ and at $B-V \approx -0.1$.  It is clear from Figure 1b that
luminosity class III stars fill in the $B-V \approx -0.1$ gap in the
luminosity class V H-R diagram, but it is not clear what 
happened to the stars in the B\"ohm-Vitense gap.  Examining the number of 
stars with luminosity classes as a function of $B-V$, we find that for 
stars preceding the gap on the blue side, 60\%-80\% have luminosity 
classes; for stars following the gap 60\% have luminosity classes; but 
for stars in the gap, only 35\%-40\% are classified.  Of the stars in the 
gap that do not have a luminosity class designation, 27\% have spectral 
type Am or Ap.

In order to demonstrate the existence of a gap, \cite{bvc74} used a
\cite{adm69} diagram (rank ordered plot of stars' $B-V$) for a flux-limited
sample of stars.  They argue
that {\it``For a narrow region of $0.10 < B-V < 0.45$ the
main-sequence stars have a small range in mass so that we may expect a
uniform distribution over masses and therefore over $T_{\rm eff}$.
Any non-uniformity in the distribution over $B-V$ can then only be due
to a nonuniform relation between $T_{\rm eff}$ and $B-V$."}  We
generate this rank diagram for luminosity class V field stars
in Figure 2 (plotted as a thin line).  With a factor of 20 more stars,
we still find a depletion in stellar density in the $0.1 < B-V < 0.4$ range 
which is consistent with the results of \cite{bvc74} and subsequent authors.

Main sequence stars might naturally be defined as
those stars to the left
of the Hertzsprung gap in the H-R diagram.  We have approximated this
division between main sequence and giants with the diagonal line in
Figure 1a.  We will refer to the set of stars with $M_V > 9.0 (B-V) - 3.3$
as the ``H-R main sequence'' to distinguish it from the
spectroscopic identification of luminosity class V stars.  When one
plots all H-R main sequence stars in the same rank diagram
(Figure 2, heavy line), then there is no depletion observed, and thus
no confirmation that there is a nonuniform distribution of $B-V$ for 
these stars.

\section {Volume-limited sample -- Houk spectral types} 

The Hipparcos data allows one to derive for the first time a 
color-magnitude diagram of a complete volume-limited sample of all stars in
our solar neighborhood out to about 100 pc and down to fluxes of about
$V<9$.  In addition, the Michigan Spectral Survey (\cite{h75}, \cite{h78}, 
\cite{h82}, and \cite{h88}) identifies spectral types and 
luminosity classifications for essentially all stars in the southern 
hemisphere ($-90 < \delta < -12$), complete to about $V< 10$.  
Our volume-limited sample consists of 8156 stars with 
Hipparcos parallax $> 10$ mas and parallax errors
typically $1\sigma = $ 1 mas (stars with errors above 30 mas are
excluded) that have been classified in the Michigan Spectral Survey.

Figure 3 shows the H-R diagram by luminosity class.  The magnitudes
used are those included with the Hipparcos Input Catalog
(\cite{gmm92}), and have photometric errors of typically 0.02 mag in
$B-V$.  Since the Houk classifications are essentially complete, we do
not have the problems with missing giants or dwarfs that were apparent in
Figure 1.  It is apparent from Figure 3 that luminosity class III draws some
A-F stars from the H-R main sequence population, 
leaving a well-defined gap in the luminosity class V H-R diagram.

Figure 4 shows the probability, as a function of $B-V$, that a star in
the H-R main sequence will be assigned a given luminosity class.
In the region $0.2 < B-V < 0.3$, only about $35\%$ of the
H-R main sequence objects are assigned to luminosity class V.  The
stars with missing classifications in the $B-V$ gap are primarily of
type Am or Ap (a few are Fm or Fp) with no luminosity class assigned.  
Only about 5\% of the luminosity class II/III, III, and III/IV stars
in the gap have m or p designations.  As noted by \cite{hm93} and 
Jasniewicz (1984), the metallic-lined A stars only partially fill the gap.

\section {Discussion and Conclusions}

The exquisite parallaxes obtained from the Hipparcos mission suggest a
resolution to puzzling gaps seen in the main sequence of field
populations and clusters.  In fact, there are no gaps in the main
sequence color-magnitude diagram afforded by Hipparcos parallaxes and
precision photometry.  Highly significant instances of the
B\"ohm-Vitense gap are only found in studies where the sample of stars
has been separated by luminosity class.  The exclusion of spectral
luminosity class III stars and stars with no luminosity class (such as
those of type Am or Ap) depletes main sequence stars with $0.2 < B-V
< 0.3$, leaving a density shortfall or gap.  Unless these excluded stars
are physically different from other main sequence stars and coincidently 
fill the gap, there is no gap caused by the sudden onset of convection.

The exact place in mass, temperature and $B-V$ color where stars
change over from a radiative to a convective atmosphere remains
undetermined from broad band photometry.
Studies of stellar chromospheric activity, which is thought to be linked to
convection, are also ambiguous (\cite{sl91}, \cite{sl97a}).  It is interesting
that \cite{r97} finds a transition in the equivalent widths of the D3
absorption feature at $B-V \approx 0.29$, indicating that this may be the
boundary between radiative and convective stars.  He notes that neither his
activity indicator nor those using C II or X-rays correlates with luminosity 
class, suggesting that the luminosity classes may have similar surface gravities.

The reason for the unexpected luminosity class identifications for
late A stars remains to be investigated.  We find it unlikely that
the majority of H-R main sequence, luminosity class III objects in the gap are 
the low surface gravity, evolved stars associated with the term
``giant" for two reasons.  First, stars in the $0.2 < B-V < 0.3$
region have notoriously weak and ambiguous spectral luminosity class 
indicators (see \cite{gg89}).  Second, the relatively large number
of these objects compared to luminosity class V stars is not consistent with 
evolutionary tracks and lifetimes of A stars.
The luminosity class III stars in Figure 3b which appear in the H-R 
main sequence may not, however, be mis-classified.  The goal of the MK
system was not to determine physical properties of stars, but rather
to present a consistent evaluation of their spectra (\cite{mk73}).
With our analysis, we cannot distinguish between misclassifications
in this region (from low signal-to-noise spectral features), and real physical
differences in the stellar spectra.  The luminosity classes in this 
region might be telling us more about subtle physical characteristics 
of the stars (such as rotation, magnetic fields, convection, chemical
composition, or binarity) than about their surface gravities or evolutionary 
stages.  The more global properties of the stars, based on broad band
photometry, allow us to place these stars on the H-R main sequence.



\acknowledgments

We acknowledge useful discussions with Roger Bell and Bob Hindsley.  We
also thank the anonymous referee and Brian Rachford for useful comments.
This work was supported by Fermi National Accelerator Laboratory, under U.S. Government
Contract No. DE-AC02-76CH03000.

\clearpage

\clearpage

\figcaption {
H-R diagram of field stars using parallax
distances from Hipparcos, high accuracy WBVR photometry, and luminosity
classes from the Hipparcos Input Catalog. 
(a) All stars in the flux-limited sample with $V<7$.  The 
sample is over $90\%$ complete.  We show the line $M_V = 9.0 (B-V) - 3.3$ for
reference.  Approximate errors
are given on the left side of the diagram.  Errors in $B-V$ are due to the
photometric errors in the WBVR catalog, which are typically 0.007 mags.
Errors in $M_V$ are primarily due to errors in parallaxes measured by
Hipparcos.  For $7{th}$ magnitude and brighter stars, the Hipparcos
satellite typically finds parallax errors of one mas.  Approximate
errors increase  for intrinsically brighter objects since they are
typically at larger distances 
in the magnitude-limited sample.  Note the smooth distribution and absence 
of gaps at $B-V\sim 0.25$; (b) Luminosity class II/III, III and 
III/IV stars;  (c) Luminosity 
class IV and IV/V stars; (d) Luminosity class V stars.  The limits of the
B\"ohm-Vitense gap ($0.22 < B-V < 0.31$), as measured by B\"ohm-Vitense and
Cantera (1974), are
shown as vertical lines.  Note that only
about $50\%$ of the stars in this sample have luminosity class assignments
in the Hipparcos catalog. \label{fig1}}

\figcaption{Cumulative distribution diagram in $B-V$ for luminosity
class V field stars from Fig. 1(d) (thin line), and those of with 
$M_V > 9.0 (B-V) - 3.3$ (thick line).  The slope of the thick line
is fairly constant over the range $0.10 < B-V < 0.45$, in contrast
to the steeper slope of the thin line in the same range. \label{fig2}}

\figcaption {
H-R diagram of field stars using parallax
distances from Hipparcos, $B-V$ from the Hipparcos Input Catalog, and luminosity
classes from the Michigan Spectral Survey. 
(a) All stars in the 
volume limited sample with $d < 100$ pc.  Typical errors are shown on the
left side of the diagram.  Again, the photometric errors contribute
very little to the errors in absolute magnitude, which are calculated
assuming one mas parallax errors at 100 pcs.  Again, the
line $M_V = 9.0 (B-V) - 3.3$ is shown for reference.
(b) Luminosity class II/III, III and III/IV stars.  (c) Luminosity class IV and IV/V stars, 
d) Luminosity class V stars.  The limits of the
B\"ohm-Vitense gap ($0.22 < B-V < 0.31$), as measured by B\"ohm-Vitense
and Canterna (1974), are
shown as vertical lines.  In contrast to the stars in Fig. 1(a), all of the
stars in Fig. 3(a) have MK classifications.  Note that luminosity class
III stars in b) tend to fill in the gap in the main sequence in d).  \label{fig3} }

\figcaption{Fraction of stars in the H-R main sequence, as defined by
$M_V > 9.0 (B-V) - 3.3$, for three sets of luminosity classes. 
We include one sigma Poisson error bars
(which are large on the blue end due to small number statistics).  Most
of the main sequence stars not included in the plot of II/III + III +
III/IV + IV + IV/V + V are metallic-lined or peculiar stars; only
a few have I or I/II designations.  For example, in the interval 
$0.175 < B-V < 0.225$, $20\%$ of the stars are luminosity
class V, $31\%$ have luminosity classes IV or IV/V, $28\%$ have luminosity
classes II/III, III, or III/IV, and $1\%$ (not shown) have luminosity classes
I or I/II.  The remaining $20\%$ are Am and Ap stars. \label{fig4}}



\clearpage


\clearpage


\clearpage


\clearpage


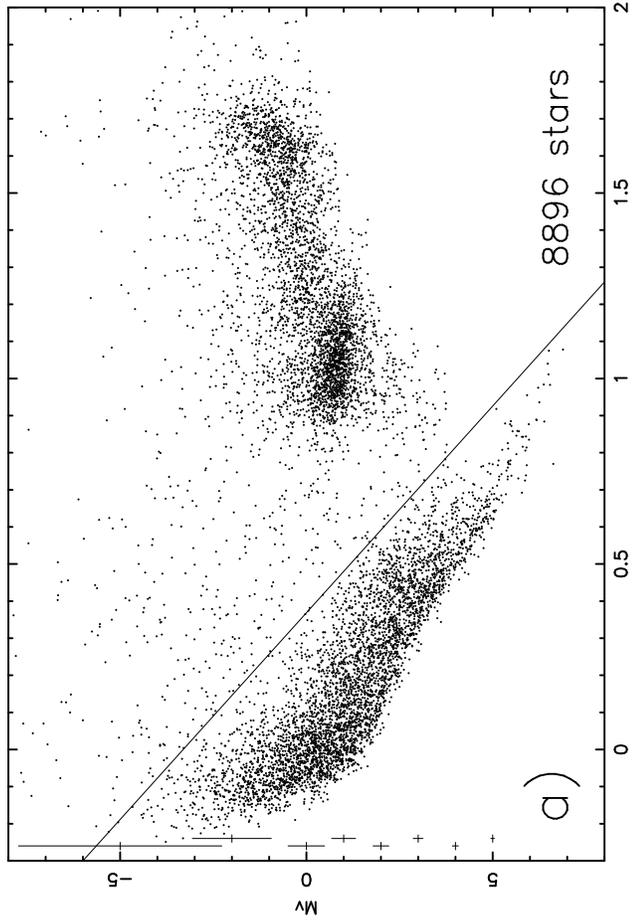
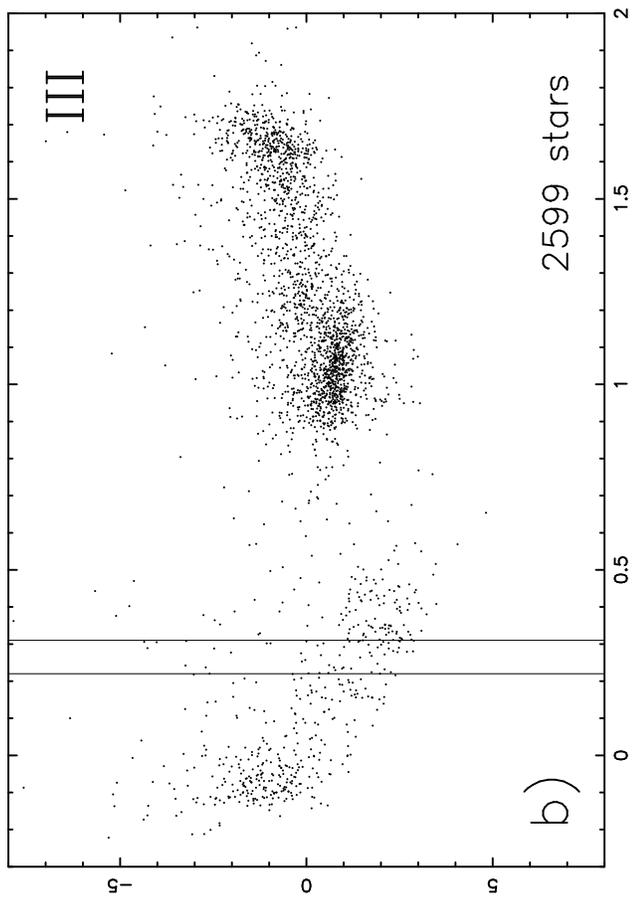
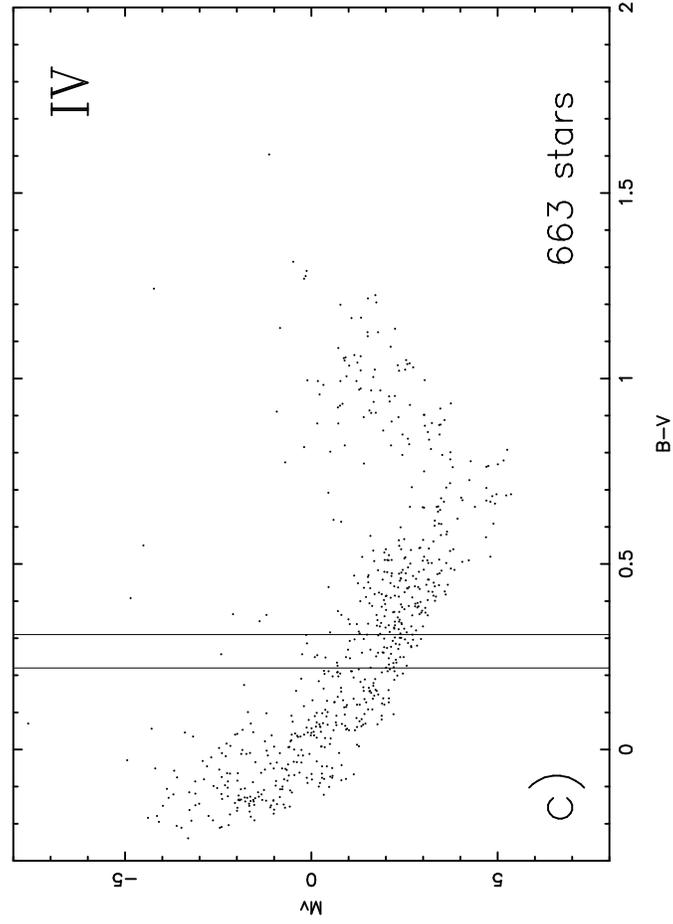
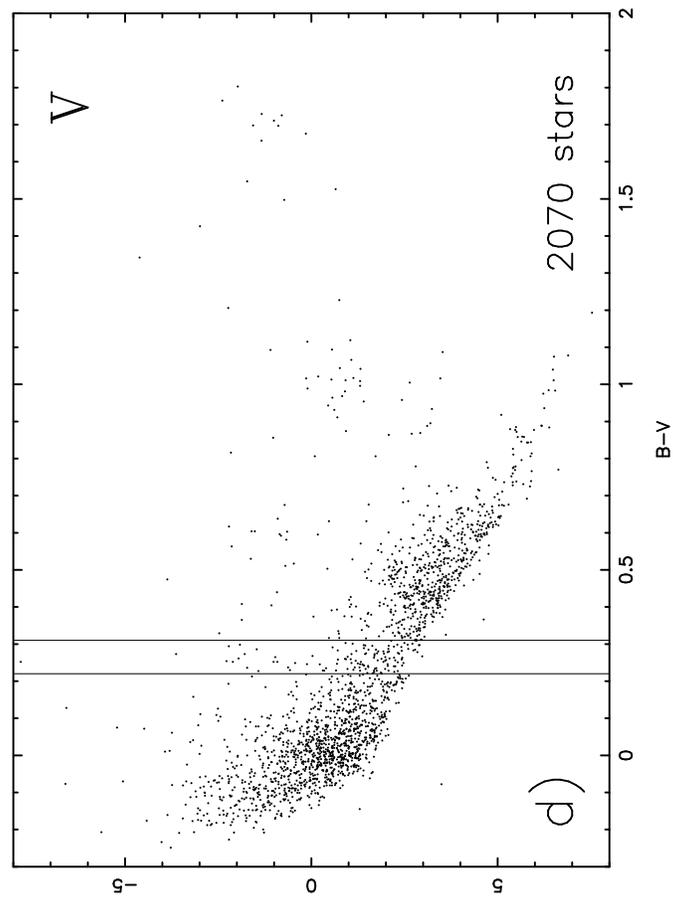
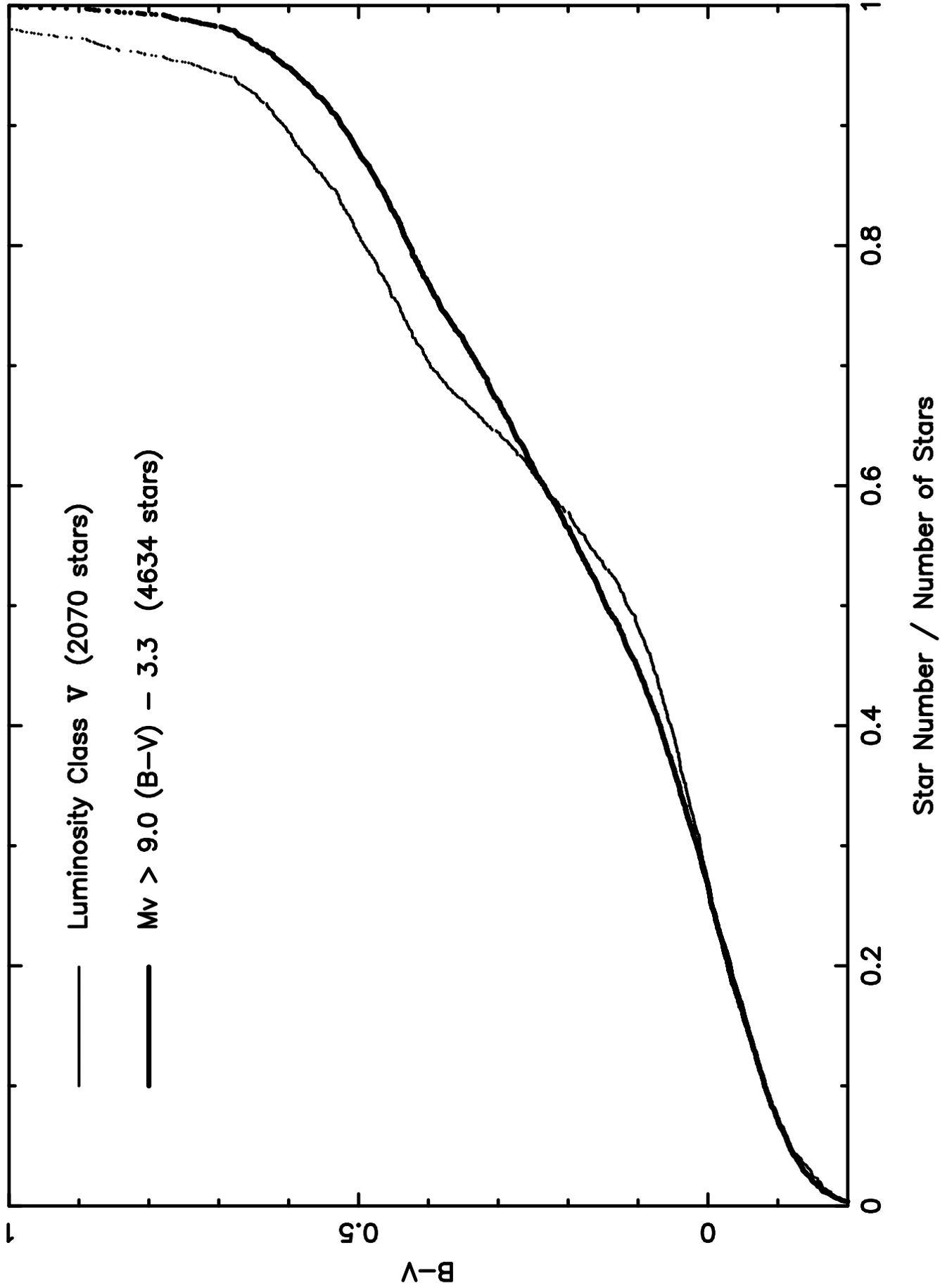
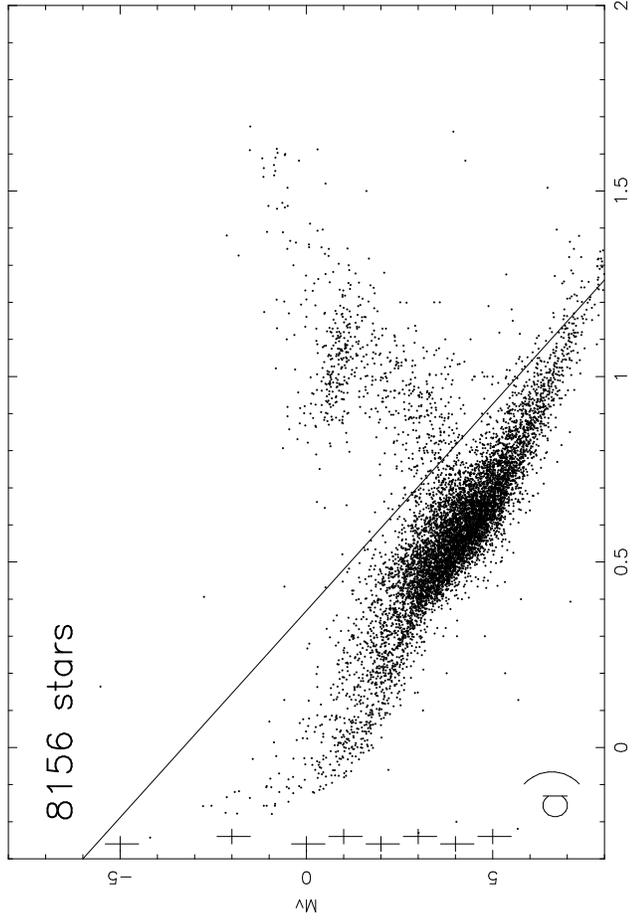
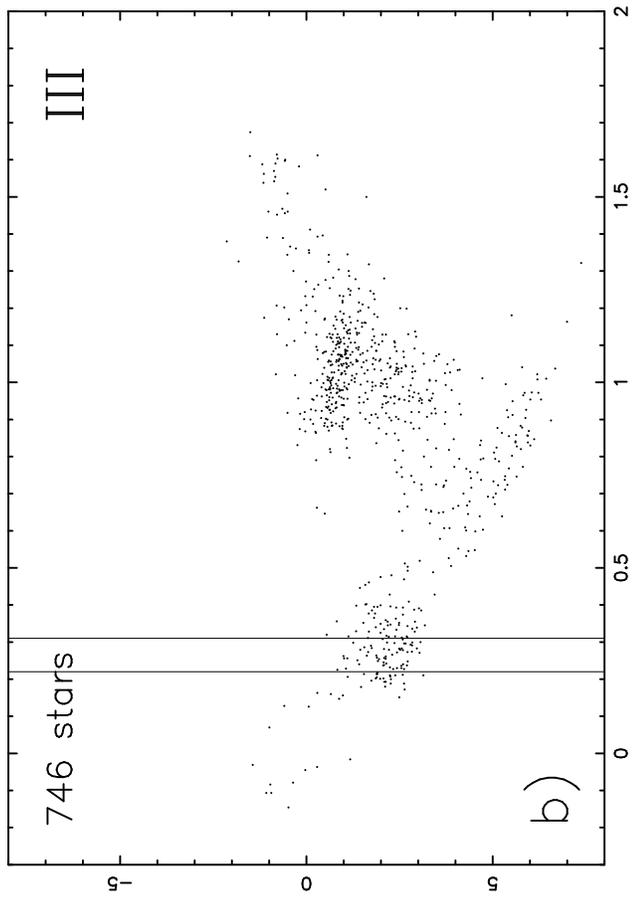
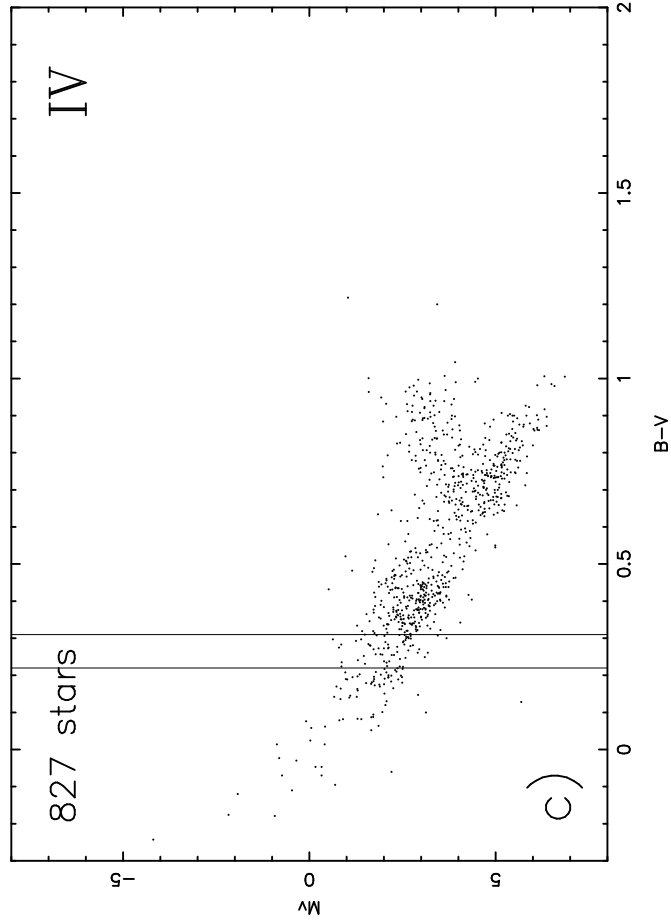
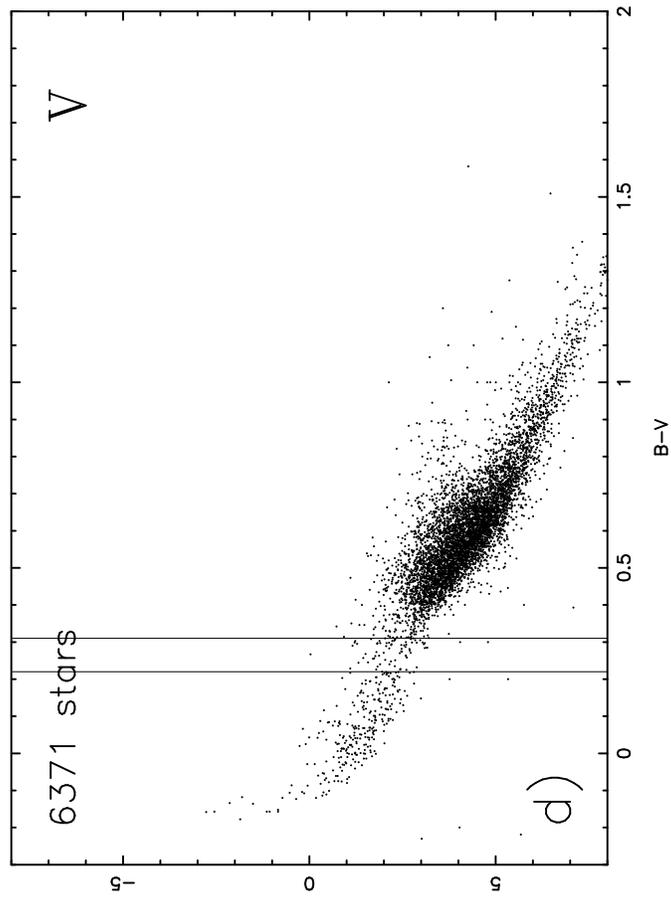
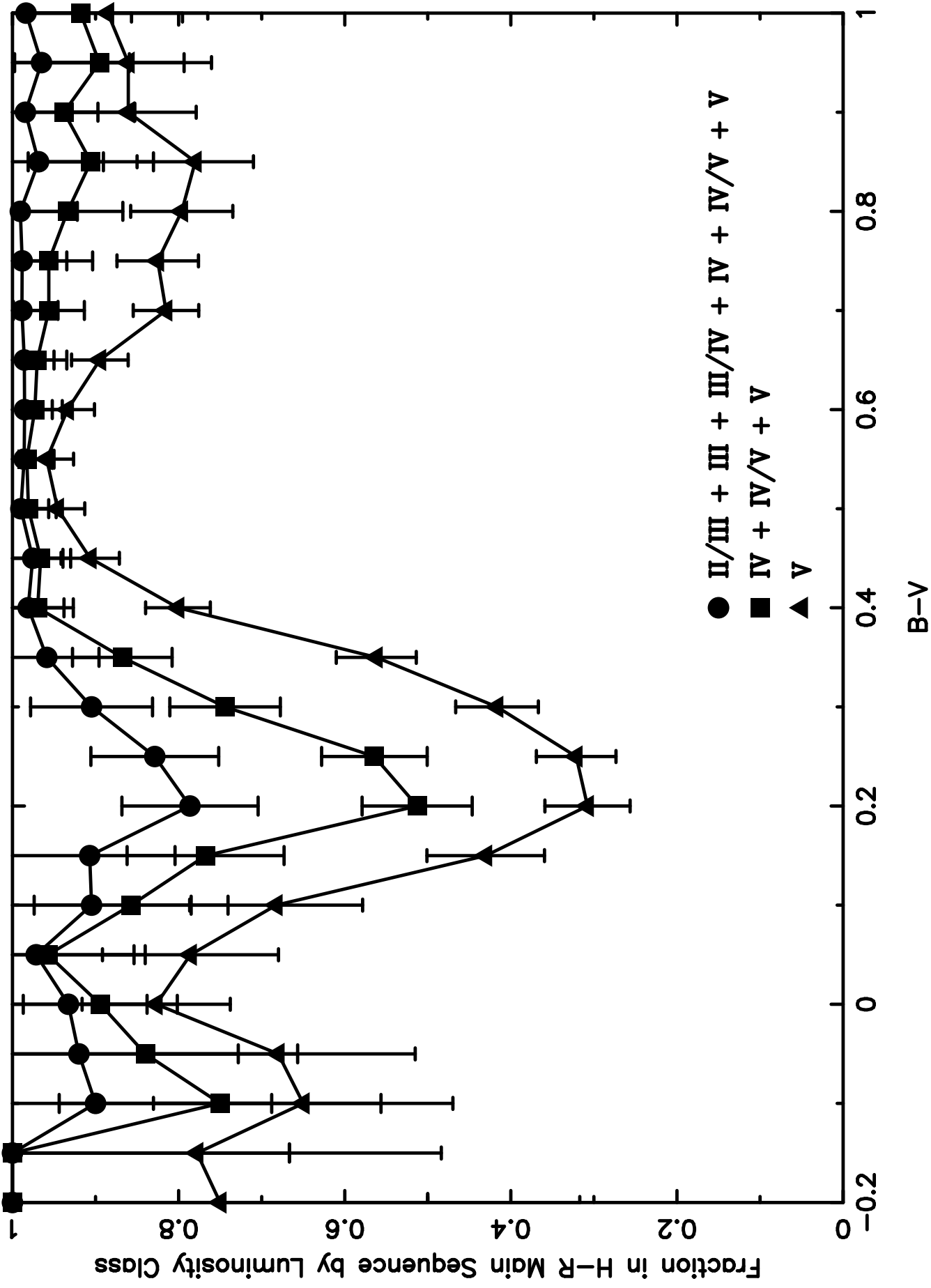
\end{document}